\documentstyle[12pt,twoside,fleqn,espcrc1,epsfig]{article}
\newcommand \ga{\raisebox{-.5ex}{$\stackrel{>}{\sim}$}}
\newcommand \la{\raisebox{-.5ex}{$\stackrel{<}{\sim}$}}

\newcommand{\AmS}{{\protect\the\textfont2
                   A\kern-.1667em\lower.5ex\hbox{M}\kern-.125emS}}
\begin{document}
\title{Dependence of lepton pair emission
       on EoS and initial state}
\author{P. Huovinen, \underline{P.V. Ruuskanen}\address{Department 
        of Physics,
        University of Jyv\"askyl\"a, Finland}
        and
        J. Sollfrank\address{Institut f\"ur Theoretische Physik,
        University of Regensburg, Germany } }

%\begin{document} % typeset front matter
\maketitle
\begin{abstract} We present results from a hydrodynamic calculation
for thermal emission of lepton pairs in central lead-lead collisions
at the CERN SPS energy. 
Dependence of the emission on the initial conditions and Equation
of State (EoS) is considered and the spectra are compared with CERES data and
calculated distribution of Drell--Yan pairs.
\end{abstract}

\section{Parametrization of initial conditions for hydrodynamics}

A major problem in using hydrodynamics to describe a high energy
nucleus-nucleus collision is how to include the primary production
stage.  At low energies hydrodynamics can be used for the description
of the whole collision process starting with the incoming nuclei as
approaching droplets of nuclear fluid which meet, compress and heat
up, followed by the expansion of this dense fireball.  At high
energies the nuclei become increasingly transparent and it becomes
unrealistic to describe the formation of the initial dense matter as a
fusion of the colliding nuclei into a single fluid droplet.  Instead,
one can parametrize the formation of matter in the form of initial
conditions for the subsequent hydrodynamic expansion.  In principle
these initial distributions should be calculated from the dynamics of
strong interactions but in practice such calculations involve
modelling, usually with several phenomenological parameters.

For our study we have adopted a simpler approach in which we
parametrize the initial state and constrain the
parametrization by experimental hadron distributions.  We base our
parametrization on the local (in transverse plane) nuclear thickness.
When combined with nuclear geometry such parametrization gives the
initial conditions for collisions of nuclei with arbitrary mass
numbers $A$ and $B$ \cite{Trento,Inicond}.  Energy and
baryon number conservation is imposed also locally in transverse plane.
With only a few
parameters we are able to reproduce the main features of hadronic
spectra in all nucleus-nucleus collisions measured at CERN
SPS \cite{Inicond}.  Even though the
parametrization is ad hoc the results contain nontrivial correlations
between different measured quantities like longitudinal and transverse
spectra and the mass dependence of the slopes of transverse spectra.
E.g., too strong stopping leads both
to too narrow rapidity distributions and too shallow transverse
distributions showing the correlation between longitudinal and
transverse flow.

In Fig.\ 1 we illustrate the mass dependence of transverse momentum
spectra on the transverse flow and the freeze-out temperature.  Lower
freeze-out temperature is reached later and leads to stronger
flow.  The net effect on the negative particles, mainly pions, which
constitute the main bulk of the matter is not large but the
heavier protons gain more from the increased common flow velocity than
what they lose in the decrease of temperature.  We observe that the
lower freeze-out temperature, $T_f\approx 120$ MeV is favoured.  For
further systematic see Ref.\ \cite{Inicond}.

Fitting the hadron spectra is not sufficient to fix the initial
conditions unambigiously.  First, some variation in stopping or
initial volume can be compensated by change of initial
densities.  Second, different equations of state (EoS)
lead to slightly different initial conditions.
For the EoSs we use, see Ref.\ \cite{Hydro}.

\begin{figure}
  \center
  \vspace{-0.7cm}
  \hspace{0.4cm}
  \begin{minipage}[t]{7.2cm}
       \epsfysize=6.5cm
       \epsfbox{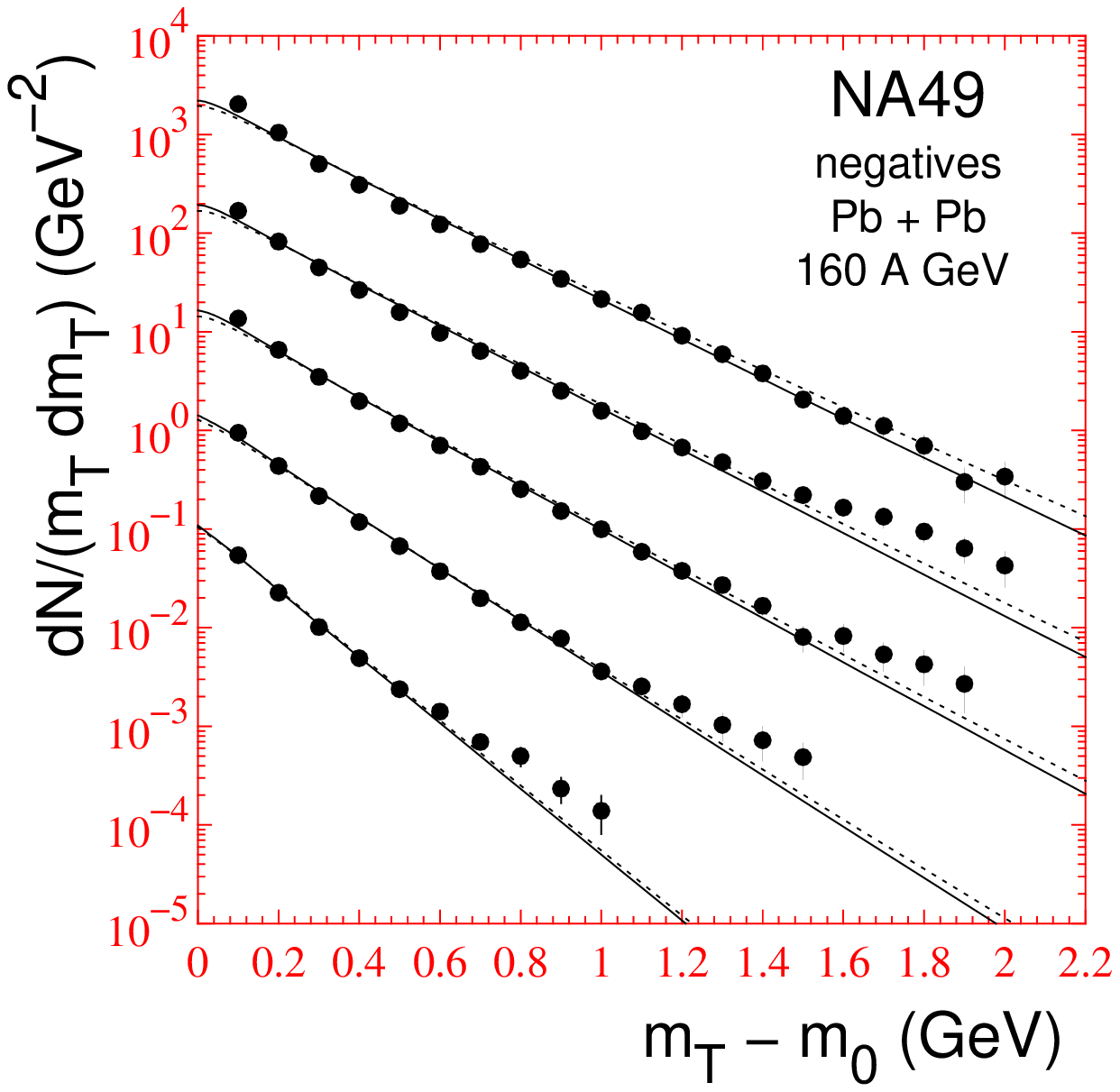}
  \end{minipage}
   \hfill
  \begin{minipage}[t]{7.3cm}
       \epsfysize=6.5cm
       \epsfbox{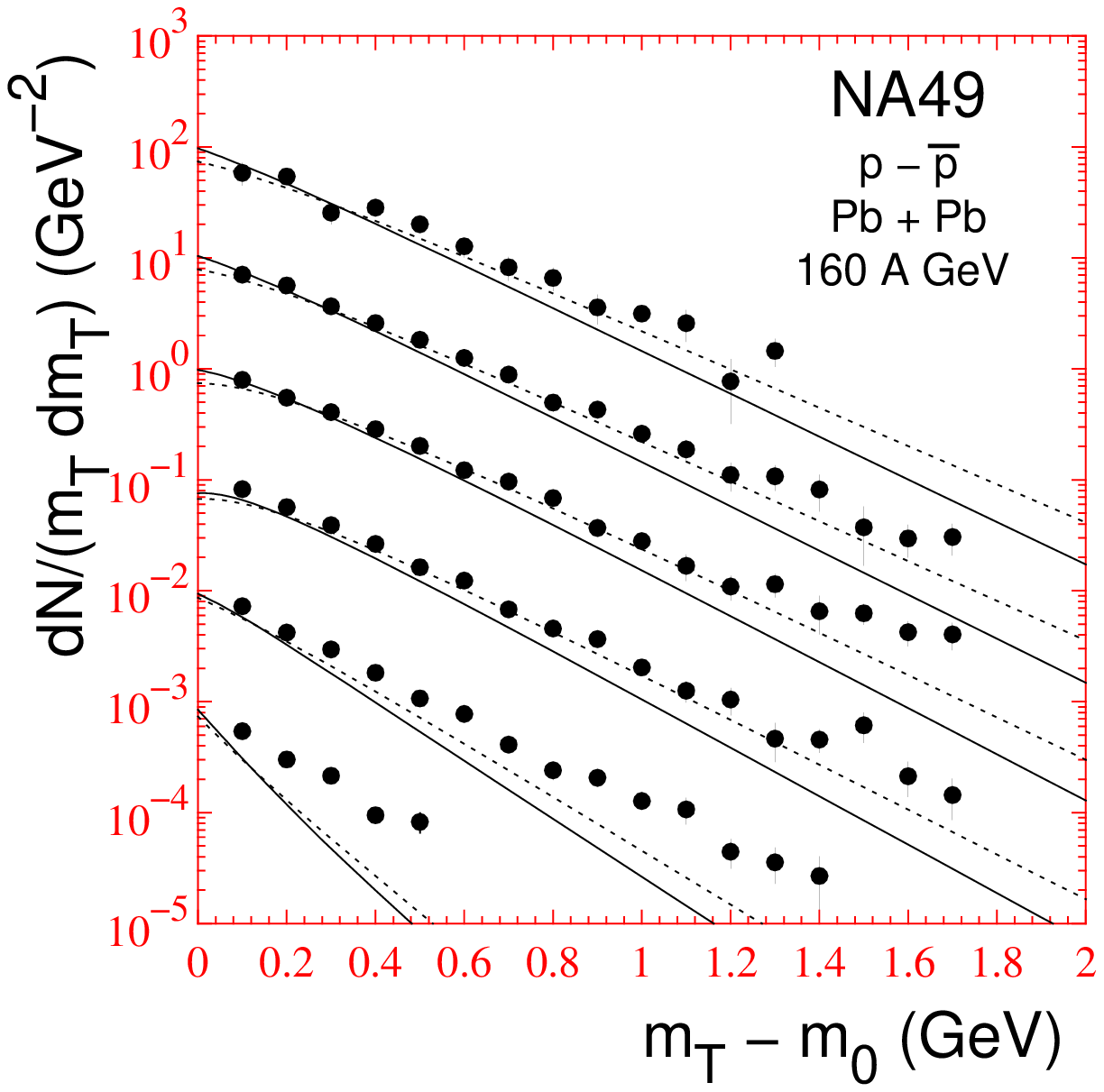}
  \end{minipage}
       \vspace{-0.8cm}
       \caption{ Transverse momentum spectra of
                negatives (left) and net protons (right) in lead-lead
                collision. Solid (dotted) line corresponds
                to $T_c\approx 140$ (120) MeV.}
       \vspace{-0.4cm}
       \label{protons}
\end{figure}

\section{Emission rates and spectra of lepton pairs}

Since the electromagnetic emission from the secondary collisions
depends on the local conditions, one can expect independent
constraints on the expansion from lepton pair and photon spectra.
Below, we compare
lepton spectra for different initial conditions and for EoSs with
transition temperatures $T_c=165$ and 200 MeV. The main difficulty are
the contributions from many background processes to these spectra.

We have studied the thermal emission of lepton pairs in the mass range
$0.3$ GeV $\le M \le 3$ GeV where excess over conventional sources has
been reported \cite{CERES,NA50,Tserr}.  We do not consider medium effects on
the $\rho$-meson parameters which are expected to be crucial for $M\le
1$ GeV. In this regime our aim is to see if the thermal contribution
is of right magnitude {\it when the evolution of the nuclear fireball
is constrained to be consistent with the observed hadron spectra}.

Our initial conditions are such that at central rapidities a large
fraction of the matter starts in the QGP phase.  For the emission
rate from the plasma we use the lowest order perturbative rate for an
ideal QGP \cite{Cleymans} . Higher order corrections are known to be important
for the emission of very low-mass pairs \cite{Pisarski,Altherr} but
the hadronic sources dominate the total spectra
in this mass region \cite{Altherr}.

In the hadron gas phase we use the binary rates of Gale and Lichard
\cite{Gale}.  In the vicinity of the transition temperature one can
argue that the thermodynamics of the hadron gas can be described by a
Hagedorn gas of noninteracting resonances.  In this
picture the strong interaction effects are embedded in the spectrum of
resonances including the vector mesons.  Assuming vector meson
dominance and quark-hadron duality, the lepton pair emission from the
resonance gas is given by the decays of vector mesons \cite{Leonidov}.
This result can be considered as an upper limit for the emission from
hadron gas when processes with extra particles in initial or final
states are neglected.  Since duality works better for heavy,
overlapping states we use this result for $M\ge 1.5$ GeV.

The excess observed by the CERES \cite{CERES} has been fitted employing
rates which take into account medium modifications but using a
simplified description of the hydrodynamic expansion \cite{Wa}
or RQMD simulation which does not admit a phase transition \cite{Ko}.
In fig.\ 2 we show our results with CERES data.
The dashed line is the background from the {\it calculated\/} final
hadrons after the freeze-out.
The full (dashed) line is the result for $T_c=200$ (165) MeV when the
thermal contribution is added.  We see that without medium
modifications the thermal emission does not fill the observed excess
of low-mass pairs even though in the $\rho$-mass region it is twice as
big as the background.  At $M\approx 500$ MeV thermal contribution is
equal to the background but the total is below the data by a factor of
4\ldots 5.  Thus an enhancement factor of $\approx$ 8\ldots 10 is
needed here if the thermal emission is the source of the excess.  The
thermal emission in the CERES region is dominated by the contribution
from hadron gas and it decreases when $T_c$ is lowered.  Since for
lower $T_c$ the hadron gas phase occurs at lower temperatures and
smaller nucleon densities, {\it the medium effects can be expected to
depend strongly on the} EoS.

In Fig.\ 3 we show the mass spectrum for $M\ga 1.5$ GeV. In the
analysis of NA50 the data is roughly a factor of 2 above the sum of
the Drell--Yan pairs and pairs from charm decays.  Since the NA50
acceptance cuts are difficult to implement \cite{Carlos} we compare
the thermal spectra with Drell-Yan pairs.  To account for the excess,
the thermal contribution should be of the same magnitude as the
Drell--Yan emission.

Because $M\gg T$ in this mass region and the emission rates are
(approximately) $\propto \exp(-M/T)$, the thermal contribution is
sensitive to the initial conditions. This is seen in Fig.\ 3 where the
dashed and dotted lines correspond to the opposite extremes of the
average initial energy density, $\langle\epsilon_i\rangle$, which
can reproduce the hadron spectra. The first conclusion
is that indeed the lepton pair contribution from the secondary
collisions can distinguish between different initial conditions.
Secondly, this contribution can be at least an important part of the
observed excess in this mass region. With our initial conditions this
region is less sensitive to the EoS.

\begin{figure}[t]
  \center
  \vspace{-0.5cm}
  \begin{minipage}[t]{7.5cm}
       \epsfxsize=6.5cm
       \hspace{0.4cm}
       \epsfbox{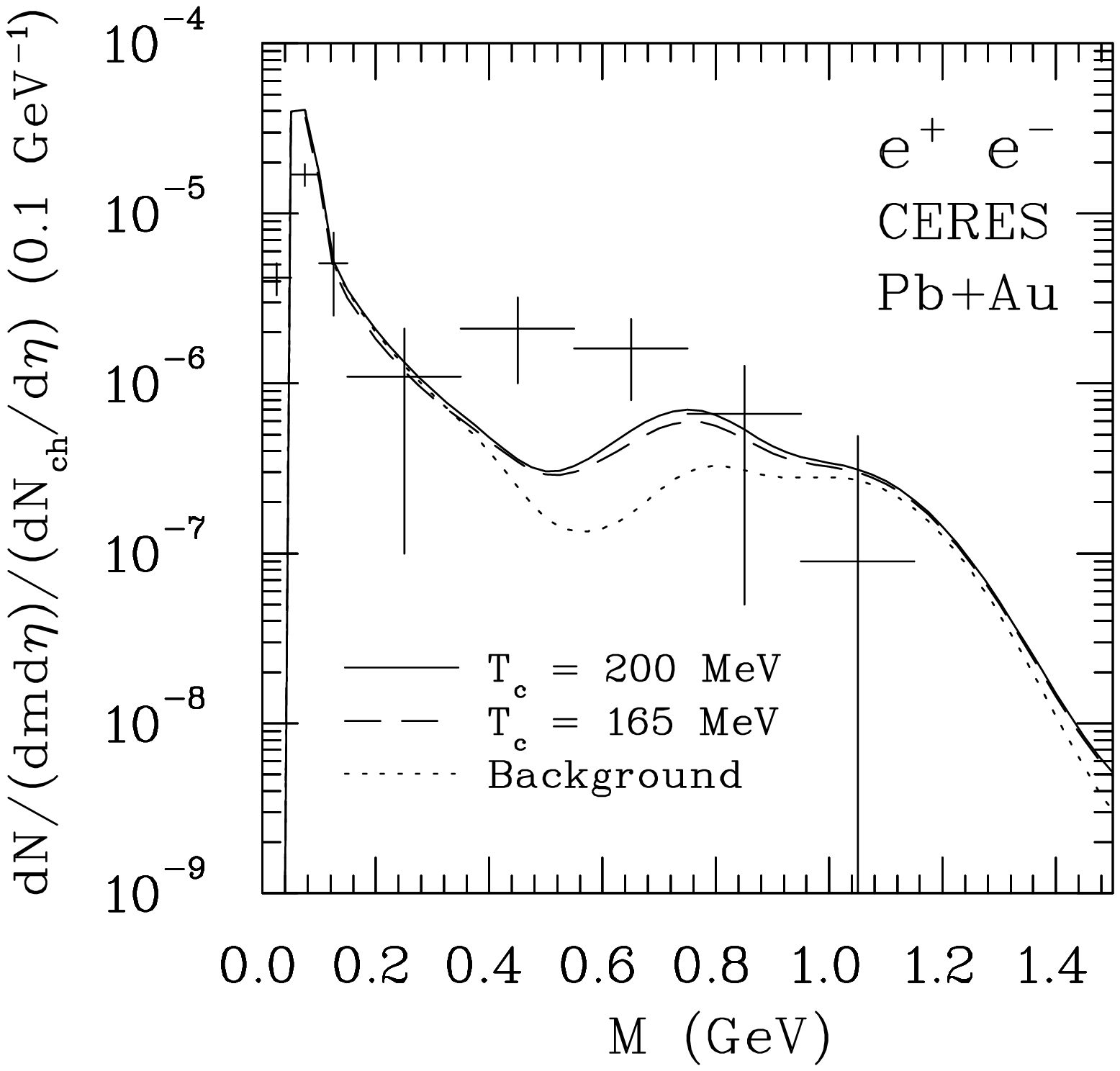}
       \vspace{-0.5cm}
       \caption{Low-mass pairs for two EoSs shown with CERES data
                \cite{CERES}.}
       \label{ceres}
  \end{minipage}
   \hfill
  \begin{minipage}[t]{7.5cm}
      \epsfxsize=6.5cm
       \hspace{0.4cm}
       \epsfbox{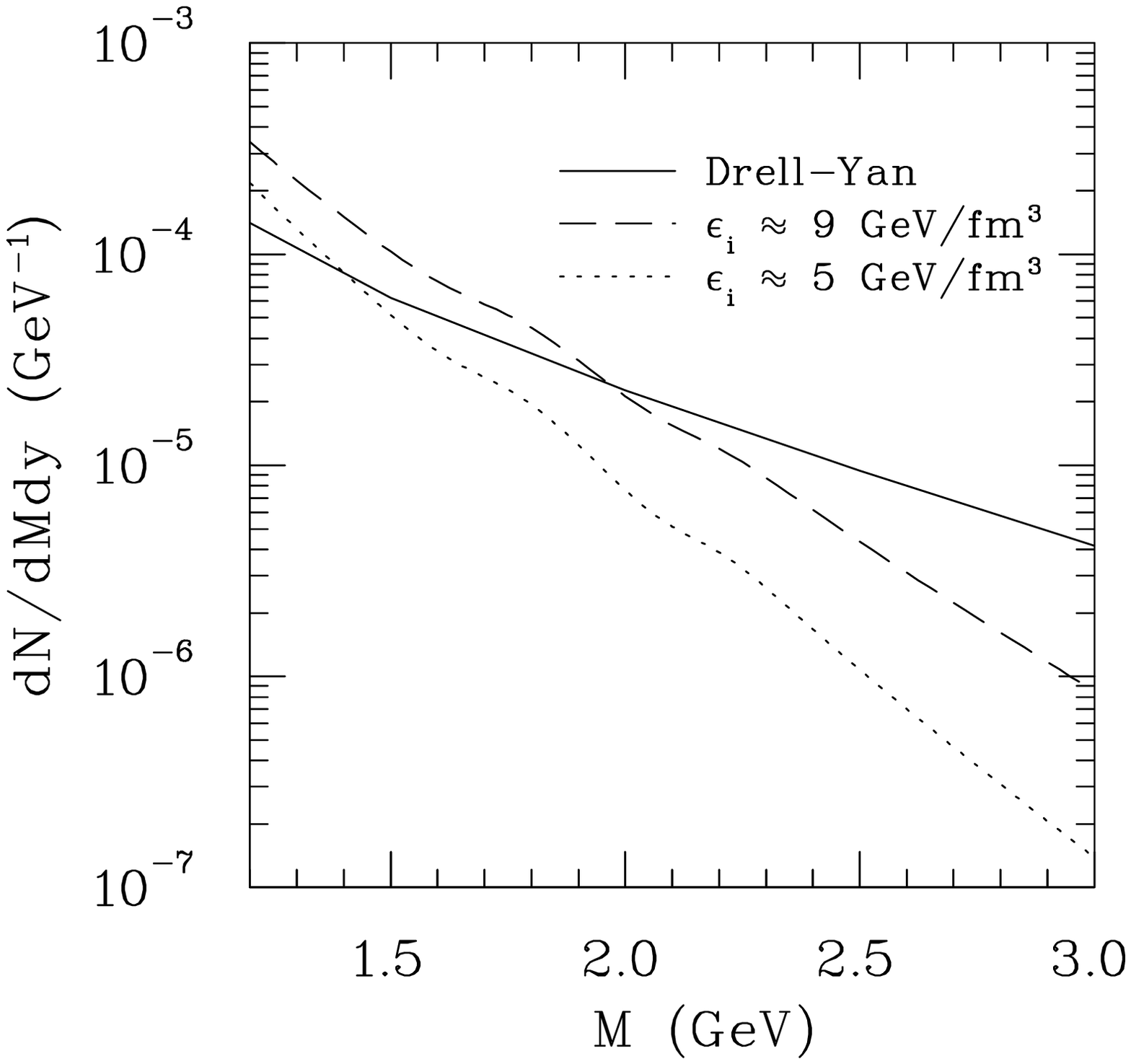}
       \vspace{-0.5cm}
       \caption{Thermal pairs for two different initial conditions
        compared with Drell--Yan pairs.}
       \label{drell}
  \end{minipage}
\end{figure}
\vspace{-0.4cm}

\section{Summary}

We have studied lepton pair emission using a simple
parametrization of initial conditions based on nuclear thickness and
conservation of energy and baryon number.  Since the parametrization
is implemented locally in the transverse plane it applies, when
combined with nuclear geometry, to collisions with arbitrary mass
numbers and can be extended to nonzero impact parameters.  It
correlates the stopping with transverse geometry.  Main features of
hadronic spectra in all nucleus-nucleus collisions measured at SPS can
be well reproduced with the same parametrization.

Hadronic observables do not completely fix the initial conditions.
Stopping and initial densities can be varied to some extent and
different choices of the EoS lead also to slightly different initial
conditions.  In the low mass region, $M\la 1$ GeV, the hadron phase
completely dominates the thermal emission and the yield is not
sensitive to the initial conditions. The yield depends on the
EoS and with the medium effects one can expect greater
sensitivity to the transition temperature $T_c$.  In the mass region,
$M\ga 1.5\ {\rm GeV}\ \gg T$ the emission is quite sensitive to the
changes in initial conditions but not on the EoS. Since the thermal
emission can be of same strength as Drell-Yan it can be an important
contribution to the excess observed by NA50 \cite{NA50}.


\begin{thebibliography}{9}

 \bibitem{Trento} J. Sollfrank, P. Huovinen and P.V. Ruuskanen, Heavy Ion
                  Physics {\bf 5}, 321 (1997).
 \bibitem{Inicond} J. Sollfrank, P. Huovinen and P.V. Ruuskanen,
                   nucl-th/9801023.
 \bibitem{Hydro} J. Sollfrank, P. Huovinen, M. Kataja, P.V. Ruuskanen,
  M. Prakash, and R. Venugopalan, Phys. Rev. C {\bf 55}, 392 (1997).
 \bibitem{CERES} P. Wurm et al., Nucl. Phys. {\bf A590} (1995) 103c;
 I. Ravinovich, these proceedings.
 \bibitem{NA50} G. Agakichiev et al., Nucl. Phys. {\bf A610} (1996) 317c.
 \bibitem{Tserr} I.\ Tserruya, these proceedings.
 \bibitem{Cleymans} J. Cleymans, J. Finberg, and K. Redlich, Phys. Rev.
 {\bf D35} (1987) 2153.
 \bibitem{Pisarski} E. Braaten, R.D. Pisarski and T.C. Yuan, Phys.
Rev. Let. {\bf 64} (1990) 2242.
 \bibitem{Altherr} T. Altherr and P.V. Ruuskanen, Nucl. Phys. {\bf
B380} (1992) 377; M.H. Thoma and C.T. Traxler, Phys. Rev.
{\bf D56} (1997) 198.
 \bibitem{Gale} C. Gale and P. Lichard,  Phys. Rev. {\bf D49} (1994) 3338;
 P. Lichard private comm.
 \bibitem{Leonidov} A. Leonidov and P.V. Ruuskanen, European Journal
of Physics, (1998), in press.
 \bibitem{Wa} J. Wambach these proceedings.
 \bibitem{Ko} C.M. Ko, G.Q. Li, G.E. Brown, and H. Sorge, Nucl. Phys. {\bf
A610} (1996) 342c; G.Q. Li, these proceedings.
 \bibitem{Carlos} C. Lourenco, private communication.

\end{thebibliography}
\end{document}